\documentclass[osajnl,twocolumn,showpacs,floatfix,bibnotes]{revtex4}

\usepackage{graphicx}

\newcommand{\be}{\begin{equation}}
\newcommand{\ee}{\end{equation}}

\newcommand\pictc[5]{\begin{figure}
                   \centerline{
                   \includegraphics[width=#1\columnwidth,height=0.7\textheight,keepaspectratio]{#3}}
               \protect\caption{\protect\label{fig:#4} #5}
                \end{figure}            }

\newcommand\pict[4][1]{\pictc{#1}{!tb}{#2}{#3}{#4}}
\newcommand\rpict[1]{\ref{fig:#1}}

\newcommand\leqt[1]{\protect\label{eq:#1}}

\newcounter{Fig}

\begin{document}
\begin{sloppy}
\title{Tunable positive and negative refraction in optically-induced photonic lattices}

\author{Christian R. Rosberg}
\author{Dragomir N. Neshev}
\author{Andrey A. Sukhorukov}
\author{Yuri S. Kivshar}

\affiliation{Nonlinear Physics Centre and Centre for
Ultrahigh-bandwidth Devices for Optical Systems (CUDOS), Research
School of Physical Sciences and Engineering, Australian National
University, Canberra, ACT 0200, Australia}
\homepage{http://www.rsphysse.anu.edu.au/nonlinear}

\author{Wieslaw Krolikowski}

\affiliation{Laser Physics Centre and Centre for
Ultrahigh-bandwidth Devices for Optical Systems (CUDOS), Research
School of Physical Sciences and Engineering, Australian National
University, Canberra, ACT 0200, Australia}

\begin{abstract}
We study tunable refraction of light in one-dimensional periodic
lattices induced optically in a photorefractive crystal. We
observe experimentally both positive and negative refraction of
beams which selectively excite the first or second spectral bands of
the periodic lattice, and demonstrate tunability of the output
beam position by dynamically adjusting the lattice depth. At
higher laser intensities, the beam broadening due to diffraction
can be suppressed through nonlinear self-focusing while preserving
the general steering properties.
\end{abstract}


\maketitle

Recent interest in the effect of negative refraction is associated
with the experimental demonstration of left-handed composite
metamaterials~\cite{Smith:2000-4184:PRL,
Parazzoli:2003-107401:PRL} which bend light in the opposite
direction to that observed in isotropic media. This negative refraction occurs due to the
effectively negative refractive index for simultaneously negative
dielectric permeability and magnetic permittivity of the medium.
However, negative refraction of waves is a fundamental physical
phenomenon that may occur in different systems as a result of
anisotropy or periodicity, and it was recently demonstrated in
two-dimensional photonic crystals~\cite{Cubukcu:2003-604:NAT}.

As a matter of fact, it has been known for many years that
negative refraction in periodic structures is possible due to the
specific properties of the extended periodic eigenmodes or Bloch
waves, and it can be observed even in weakly modulated
one-dimensional periodic
lattices~\cite{Russell:1995-585:ConfinedElectrons}. Indeed, when
light bends at the interface between homogeneous and periodic media, the refraction
angle depends on the effective diffraction coefficients of the particular Bloch
waves of the structure. Since diffraction of the Bloch waves depends strongly on
the refractive index contrast, the refraction angle can be
controlled by dynamically varying the lattice depth. Such tunability of the lattice depth and negative refraction
can be achieved in several physical systems, including
holographic gratings induced optically in photorefractive
crystals~\cite{Efremidis:2002-46602:PRE, Fleischer:2003-23902:PRL,
Neshev:2003-710:OL} and liquid-crystal waveguides with patterned
electrodes~\cite{Fratalocchi:2005-174:OL}.

In this Letter, we study theoretically and demonstrate
experimentally the control over light refraction in
optically-induced photonic lattices. For the first time to our
knowledge, we realize dynamic tunability of the beam refraction
associated with different Bloch waves by using a tilted lattice of
a variable depth. We demonstrate that while beams corresponding to
the first band of the lattice bandgap spectrum are positively refracted following
the direction of the lattice, the beams associated with the
top of the second band experience enhanced negative refraction in the direction opposite to the lattice tilt.
Furthermore, we combine such tunability with spatial localization
of the beams through nonlinear self-focusing in the normal
diffraction regime.

We consider an optically-induced lattice in a biased photorefractive
crystal where the propagation of extraordinarily polarized beams with amplitude $E(x,z)$ is
described by the equation
\be \leqt{nls}
   i \frac{\partial E}{\partial z}
   + D \frac{\partial^2 E}{\partial x^2}
   + {\cal F}( x-\alpha z, |E|^2) E   = 0 ,
\ee
where $x$ and $z$ are the transverse and propagation coordinates
normalized to the characteristic values $x_0$ and $z_0$,
respectively, $D = z_0 \lambda / (4 \pi n_0 x_0^2)$ is the beam
diffraction coefficient, $\alpha$ is the normalized angle of the
lattice tilt [see Fig.~\rpict{theory}(a)], $n_0$ is the average
refractive index of the medium, $\lambda$ is the wavelength in vacuum,
${\cal F}(x,|E|^2) = - \gamma \left[I_b + I_g \cos^2(\pi x / d) +
|E|^2\right]^{-1}$, $I_b$ is the constant dark irradiance, $I_g$
is the peak intensity of the interference pattern of period $d$,
and $\gamma$ is a nonlinear coefficient linear proportional to the
applied DC field. To match our experimental conditions, we use the
following parameters: $\lambda = 0.532~\mu$m, $n_0 = 2.4$, $x_0 = 1~\mu$m, $z_0 = 1$mm, $d = 19.2$, $I_b =1$, $I_g = 1$, and the
crystal length $L = 15$~mm. Then, the refractive index contrast in the lattice is
$\Delta n = \gamma \lambda / (4 \pi z_0)$.

Propagation of linear waves in a {\em straight lattice}
($\alpha=0$) is defined through the spectrum of Bloch
waves~\cite{Neshev:2004-83905:PRL}, $E(x,z) = \psi(x) \exp(i K x/d
+ i \beta z)$, where $\psi(x)$ is periodic, $K$ is the Bloch wave
number, and $\beta$ is the propagation constant. Dispersion curves
$\beta(K)$ form bands, as shown in Figs.~\rpict{theory}(b,c). In a
{\em tilted lattice}, the Bloch-wave dispersion can be written as
$\widetilde{\beta}[K] = \beta[K-\alpha/(2D)] - K \alpha + \alpha^2
/ (4 D)$, which indicates that the dispersion curves are
translated and tilted simultaneously with $\alpha$.

When an input beam excites Bloch waves from a particular spectral
band, its normalized propagation angle inside the lattice can be
found as $\theta = -d\widetilde{\beta}/dK$. Waves corresponding to
the middle ($K=0$) or edge ($K=\pi$) of the Brillouin zone always
propagate straight if the lattice is not tilted ($\alpha=0$),
since $\theta=0$ at all band edges [see
Figs.~\rpict{theory}(b,c)]. However, in a tilted lattice the same waves
refract at an angle $\theta = -
\partial \beta[K-\alpha/(2D)] / \partial K + \alpha \simeq \alpha
(D - D_{\rm eff}) / D + O(\alpha^2)$, where $D_{\rm eff} =- (1/2)
d^2 \beta / d K^2$ is the effective diffraction coefficient of
Bloch waves. Thus, the beam refraction induced by the lattice tilt
is proportional to the difference of the diffraction coefficients
in a bulk crystal and in the lattice at the corresponding $K$.

Results of our calculations presented in Fig.~\rpict{theory}(d)
show that for the top of the first band $0 < D_{\rm eff} < D$, and
therefore the beam is refracted in the direction of the lattice
tilt [Figs.~\rpict{theory}(e,f), top]. At the bottom of the first
band, $D_{\rm eff} < 0$, and the beam is refracted even stronger
in the same direction. In a sharp contrast, for the top of the
second band $D_{\rm eff} > D$ and the beam experiences {\em
negative refraction} [Fig.~\rpict{theory}(e), bottom] in shallow
lattices with $\Delta n < \Delta n_{\rm cr}$, whereas $D_{\rm eff} <
D$ and {\em positive refraction} should occur in deeper lattices
with $\Delta n > \Delta n_{\rm cr}$. Most remarkably, in the critical
case when $\Delta n \simeq \Delta n_{\rm cr}$ and $D_{\rm eff} \simeq
D$, the band-2 beam always goes straight [Fig.~\rpict{theory}(f),
bottom] even when the lattice is tilted (at relatively small angles).
Linear diffraction leads to beam broadening which can be
suppressed in nonlinear media due to the effect of self-focusing.
In the examples shown in Figs.~\rpict{theory}(e,f), we have chosen
the wave intensity to achieve the formation of {\em lattice
solitons} which do not diffract and have constant width. These
mobile solitons spanning several lattice periods exhibit the same
refraction as linear beams.

\pict{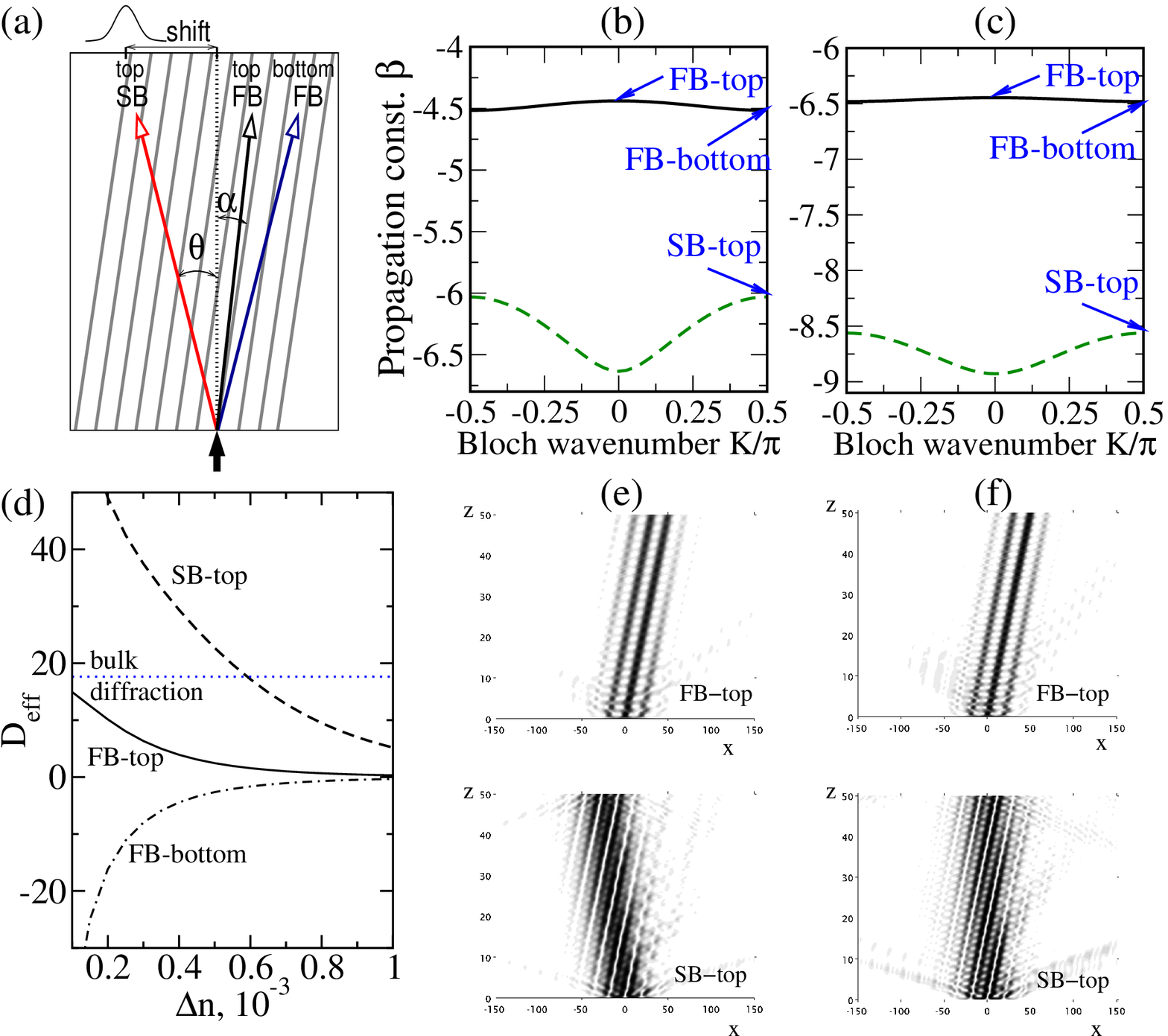}{theory}{ (a) Schematic of the beam refraction
associated with different Bloch modes: at the top and bottom of
the first band (FB) and top of second band (SB). 
(b,c)~Bloch-wave dispersion curves in (b)~shallow ($\Delta n = 0.3\times10^{-3}$) and (c)~deeper ($\Delta n = 0.45\times10^{-3}$) lattices. 
(d)~Effective diffraction coefficients at the band edges vs. the lattice index contrast $\Delta n$ compared with the bulk diffraction $D$.
(e,f)~Numerical results for refraction of beams associated with the top of the first
and second bands in lattices corresponding to~(b,c).
}

In experiment, we induce a periodic lattice in a
$15\times5\times5$~mm SBN:60 photorefractive crystal externally
biased and homogeneously illuminated with white
light~\cite{Fleischer:2003-23902:PRL,Neshev:2003-710:OL}. An
optical lattice with the period of 19.2~$\mu$m is created by
interfering two ordinarily-polarized broad beams from a
frequency-doubled Nd:YVO$_4$ laser at 532~nm. By varying the bias
voltage, we change the contrast of the refractive index
modulation of the lattice, thus altering the bandgap structure and
correspondingly modifying the diffraction properties of the Bloch
waves.

\pict{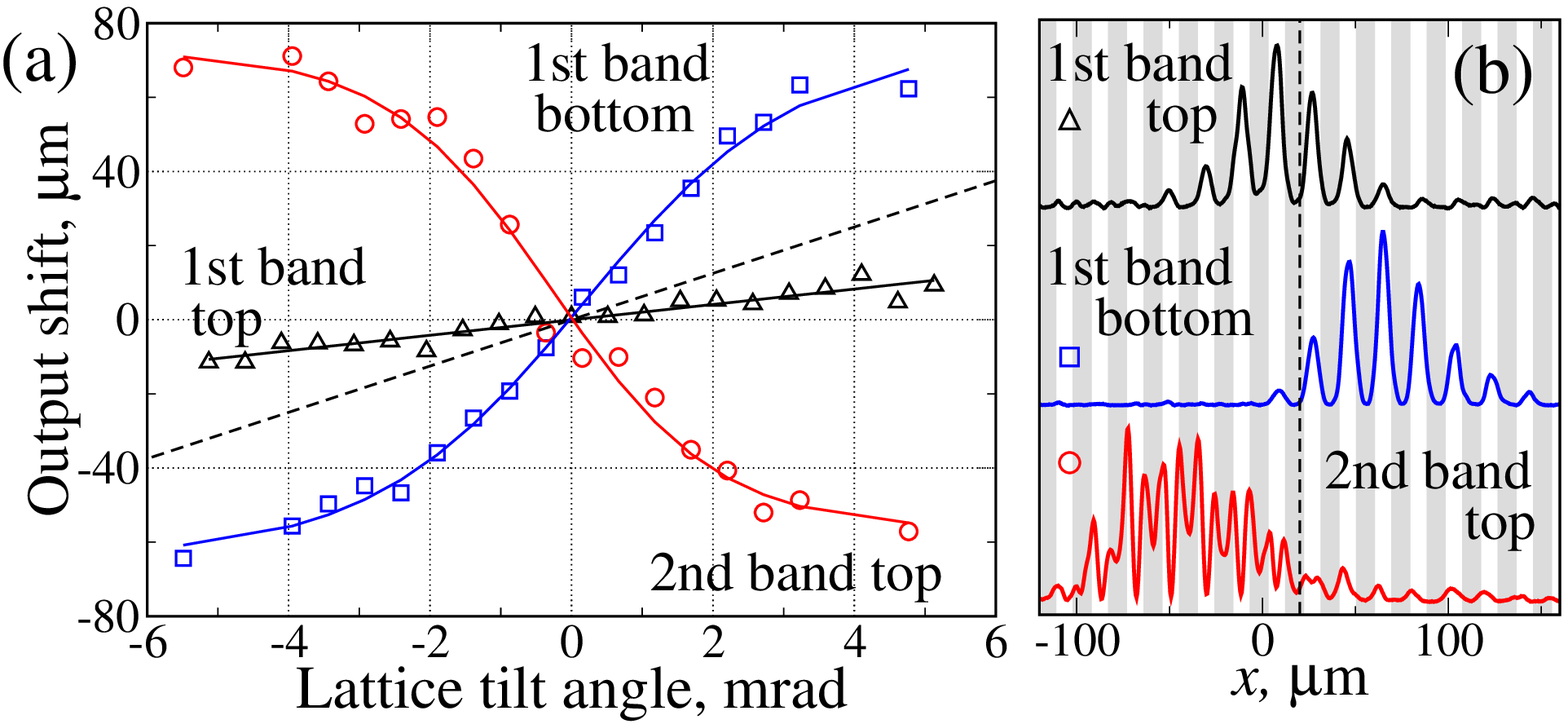}{angle_tunability}{ (a) Shift of the output
position of beams associated with three different Bloch waves
vs. the lattice tilt (measured in air). (b) Characteristic beam
profiles for a tilt of 3~mrad. In (a,b) the dashed line corresponds
to the shift of the lattice. Bias voltage is 1.5~kV. }

Different Bloch waves of the induced periodic structure are
excited by several extraordinarily polarized beams. The wave at the top of the
first band is excited by a single Gaussian beam of a full width at
half maximum (FWHM) of 30~$\mu$m, propagating along the induced
lattice~\cite{Neshev:2003-710:OL}. To selectively excite Bloch
waves at the two edges of the Bragg-reflection gap, we
employ a two-beam excitation
technique~\cite{Neshev:2004-83905:PRL}, where a pair of beams, each
propagating under the Bragg angle (in opposite directions), is
focused at the front face of the crystal to a FWHM of 65~$\mu$m.
By tilting the lattice without changing the initial direction of
the probe beams [Fig.~\rpict{theory}(a)], we study the steering of
the propagating beams. This steering strongly depends on the
lattice tilt, as the latter changes the diffraction coefficient of the
Bloch waves by scanning along the dispersion curves
[Fig.~\rpict{theory}(b,c)]. In Fig.~\rpict{angle_tunability}(a) we
show the change of the beam position at the output face of the
crystal as a function of the lattice tilt. The refraction is
positive for waves in the first band and negative
for those in the second band. Consequently, beams associated with these two bands are seen to be shifted in opposite directions. For the two waves at the edges of the Bragg reflection gap, the output shift approaches that of a beam
propagating at the Bragg angle (corresponding shift of 76~$\mu$m) as the lattice tilt is increased. This effect can be considered as an angular amplification of beam
deflection since the output propagation angle is substantially larger than the initial lattice tilt.
In our particular case, we measure a {\em six fold} increase of the
tilt angle with a positive or negative gain for the waves of
the top and the bottom of the spectral gap, respectively. On the other hand, the deflection of waves from the top of the first band is much smaller and less sensitive to the lattice tilt. The corresponding output beam profiles for a lattice tilt of 3~mrad are depicted in
Fig.~\rpict{angle_tunability}(b). It is clearly seen that these profiles are well represented by the structure of the
corresponding Bloch waves superimposed on a bell-shaped envelope. 
These findings are in full agreement
with our theoretical predictions.

\pict{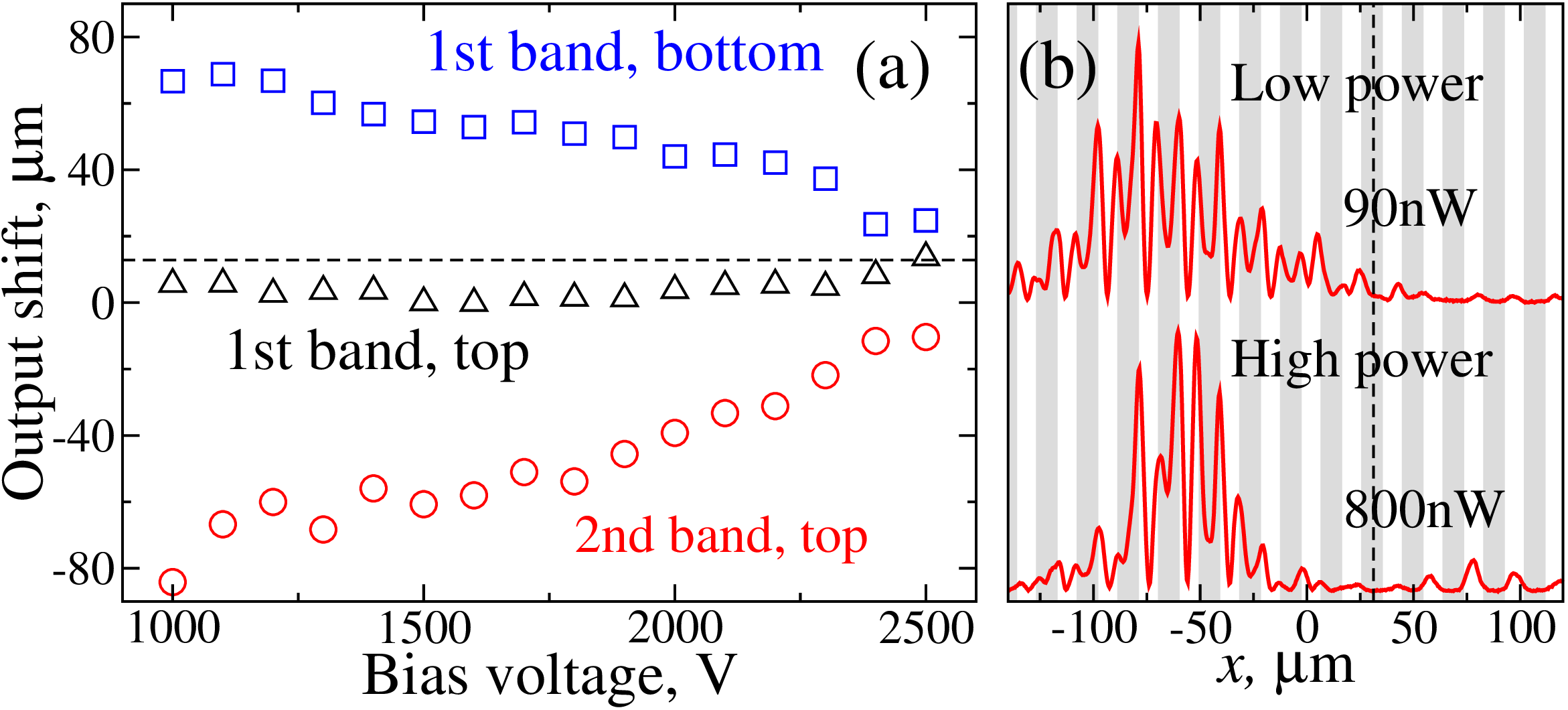}{voltage_tunability}{ (a) Measured output shift
vs. bias voltage at low power for beams corresponding to three
different Bloch modes. (b) Output beam intensity profiles (at
1.8~kV) corresponding to the top of the second band for low (top)
and high (bottom) power. In (a,b) the dashed line corresponds to the
lattice shift, and the lattice tilt is 2~mrad.}

In order to demonstrate tunability of the beam refraction and
steering, we measured the change of the beam position at the crystal output face
as a function of bias voltage.
In Fig.~\rpict{voltage_tunability}(a) we show the output beam
shift for the three different Bloch modes vs. the applied voltage. The lattice is tilted
to the right at an angle of 2~mrad, measured in air, corresponding to 15\% of the Bragg angle. At low voltages (shallow
lattice) waves corresponding to the bottom of the first band and the top of the second band experience positive and negative
refraction, respectively. However, increasing the bias
voltage results in reduced beam mobility and hence a monotonical decrease of the output shift for both beams. The beam deflection can be almost entirely
suppressed for a deep lattice, corresponding to voltages higher
than 2.5~kV. For voltages lower than 1000~V, the lattice is too shallow to support
beams of the particular spectral width. In contrast to this strongly voltage dependent steering of beams belonging to the edges of the Bragg reflection gap, the beam associated with the top of the first band is seen to be deflected less
than the lattice itself, and this deflection does not
change significantly with the voltage.

An important requirement for practical realization of controllable beam
steering is that beams stay localized as they propagate through
the sample. For a focusing nonlinearity, such
localization can be realized for beams at the top of
the bands, where diffraction is positive. In this case the propagation constant
moves inside the gap at increased laser intensity and
leads to formation of lattice solitons. An important
question, however, is whether the effect of localization preserves the
steering properties of Bloch waves. In
Fig.~\rpict{voltage_tunability}(b) we show two intensity profiles
of a beam associated with the top of the second band, taken at low
(90~nW) and high power (800~nW). At higher laser power
the beam is self focused from an output FWHM of 81~$\mu$m in the linear regime to 46
$\mu$m in the nonlinear regime. It is seen that the effect of self-focusing preserves the
negative refraction properties, with only a small decrease of the
deflection angle (10~$\mu$m at the output). We would like to point out that the
degree of self-focusing depends on the lattice tilt, since this
tilt determines the effective beam diffraction.

In conclusion, we have studied theoretically and demonstrated
experimentally tunable refraction of light in optically-induced
one-dimensional photonic lattices. We have observed negative
refraction associated with the top of the second band of the
lattice bandgap spectrum, and demonstrated tunability of the
output beam position by a dynamically reconfigurable lattice
depth. We have studied the light propagation at higher laser
intensities for self-focusing nonlinearity and observed that the
basic steering properties and negative refraction are preserved in
the self-trapping regime.

The authors thank Ben Eggleton and Martjin de Sterke for useful
discussions and acknowledge support from the Australian Research
Council.

\end{sloppy}

\begin{thebibliography}{1}

\bibitem{Smith:2000-4184:PRL}
D.~R. Smith, W.~J. Padilla, D.~C. Vier, S.~C. Nemat~Nasser, and S. Schultz,
  Phys. Rev. Lett. {\bf 84}, 4184 (2000).

\bibitem{Parazzoli:2003-107401:PRL}
C.~G. Parazzoli, R.~B. Greegor, K. Li, B.~E.~C. Koltenbah, and M. Tanielian,
  Phys. Rev. Lett. {\bf 90}, 107401 (2003).

\bibitem{Cubukcu:2003-604:NAT}
E. Cubukcu, K. Aydin, E. Ozbay, S. Foteinopoulou, and C.~M. Soukoulis, Nature
  {\bf 423}, 604 (2003).

\bibitem{Russell:1995-585:ConfinedElectrons}
P.~St.~J. Russell, T.~A. Birks, and F.~D. Lloyd~Lucas, ``Photonic Bloch waves
  and photonic band gaps,''  in {\em Confined Electrons and Photons}, E.
  Burstein and C. Weisbuch, eds., (1995), \ pp.\ 585--633.

\bibitem{Fleischer:2003-23902:PRL}
J.~W. Fleischer, T. Carmon, M. Segev, N.~K. Efremidis, and D.~N.
  Christodoulides, Phys. Rev. Lett. {\bf 90}, 023902 (2003).

\bibitem{Neshev:2003-710:OL}
D. Neshev, E. Ostrovskaya, Y. Kivshar, and W. Krolikowski, Opt. Lett. {\bf 28},
  710 (2003).

\bibitem{Efremidis:2002-46602:PRE}
N.~K. Efremidis, S. Sears, D.~N. Christodoulides, J.~W. Fleischer, and M.
  Segev, Phys. Rev. E {\bf 66}, 046602 (2002).

\bibitem{Fratalocchi:2005-174:OL}
A. Fratalocchi, G. Assanto, K.~A. Brzdakiewicz, and M.~A. Karpierz, Opt. Lett.
  {\bf 30}, 174 (2005).

\bibitem{Neshev:2004-83905:PRL}
D. Neshev, A.~A. Sukhorukov, B. Hanna, W. Krolikowski, and Yu.~S. Kivshar,
  Phys. Rev. Lett. {\bf 93}, 083905 (2004).

\end{thebibliography}
\end{document}